\documentstyle[12pt]{article}
\newcommand{\pslash}{\not \! p}

\begin{document}

\begin{flushright}{UT02-47 }
\end{flushright}
\vskip 0.5 truecm

\begin{center}
{\Large{\bf N=2 Wess-Zumino model on the d=2 Euclidean 
lattice}}
\end{center}
\vskip .5 truecm
\centerline{\bf Kazuo Fujikawa }
\vskip .4 truecm
\centerline {\it Department of Physics,University of Tokyo}
\centerline {\it Bunkyo-ku,Tokyo 113,Japan}
\vskip 0.5 truecm

\makeatletter
\@addtoreset{equation}{section}
\def\theequation{\thesection.\arabic{equation}}
\makeatother

\begin{abstract}
We examine the $N=2$ Wess-Zumino model defined on the 
$d=2$ Euclidean lattice in connection with a restoration of 
the Leibniz rule in the limit $a\rightarrow0$ in perturbatively
finite theory. We are interested in 
the difference between the Wilson and Ginsparg-Wilson
fermions and in the effects of extra 
interactions introduced by an analysis of Nicolai mapping. 
As for the Wilson fermion, it induces a linear divergence  to 
individual tadpole diagrams in the limit $a\rightarrow0$, which 
is absent in the 
Ginsparg-Wilson fermion. This divergence suggests that a
careful choice of lattice regularization is 
required in a reliable numerical simulation. 
As for the effects of the extra couplings introduced by an 
analysis 
of Nicolai mapping, the extra couplings do not completely remedy
 the supersymmetry breaking in correlation functions induced 
by the failure of the Leibniz rule in perturbation 
theory, though those couplings ensure the 
vanishing of vacuum energy arising from disconnected diagrams.
Supersymmetry in correlation functions is recovered
in the limit $a\rightarrow 0$ {\em with or without} those 
extra couplings.  
In the context of lattice theory, it may be properly said that
 supersymmetry does not improve ultraviolet properties but 
rather it is well accommodated in theories with good 
ultraviolet properties.
\end{abstract}

\large

\section{Introduction}
It is important to define the supersymmetric Wess-Zumino 
model\cite{wess}
on the lattice in such a way that the non-renormalization 
theorem\cite{iliopoulos}\cite{fujikawa} and consequently the 
absence of  quadratic
divergence is preserved. Since the absence of  quadratic
divergence arises from a subtle cancellation of bosonic
and fermionic contributions, we have to ensure the 
precise (not approximate) bose-fermi cancellation. This task is
not easy if one recalls that the Leibniz rule is generally 
broken on the lattice\cite{dondi}. See Refs.\cite{nicolai}-
\cite{fuji-ishi} for the analyses of related issues.  

Recently, it was suggested\cite{fujikawa2} that a 
perturbatively finite 
theory, if latticized, could preserve  supersymmetry
to all orders in perturbation theory in the sense that 
the supersymmetry breaking terms induced by the failure of 
the Leibniz rule become irrelevant in the limit $a\rightarrow0$.
It was demonstrated in Ref.\cite{fujikawa2} that this is in 
fact realized if one first renders the 4-dimensional 
Wess-Zumino model finite by applying the higher derivative
regularization. A non-perturbative confirmation of this proposal
has not been given yet, but we believe that a perturbative
confirmation of the absence of quadratic divergence 
is a prerequisite for the non-perturbative analysis\footnote{It 
should be noted that the lattice in this context is introduced 
not to control the divergences but to make numerical and other 
non-perturbative analyses possible.}. 
In Ref.\cite{fujikawa2}, the Ginsparg-Wilson 
fermion\cite{neuberger}\cite{fujikawa3}\cite{niedermayer} was 
utilized, which 
has a nice chiral property but at the same time introduces 
certain subtle aspects to the analysis\cite{fuji-ishi}. 

A 2-dimensional reduction of the Wess-Zumino model,
which exhibits $N=2$ supersymmetry, is finite perturbatively
with qualifications to be specified later,
and thus it provides a good testing ground of the suggestion 
made in Ref.\cite{fujikawa2}, though the crucial issue of 
quadratic divergence cannot be studied in this model. In the 
present paper, we examine 
the $N=2$ 
Wess-Zumino model on the $d=2$ Euclidean lattice  
perturbatively and clarify several basic issues involved which 
may become relevant in an actual numerical simulation. 
Firstly, we examine the 
use of the Wilson fermion instead of the Ginsparg-Wilson fermion
since the Wilson fermion is much easier to handle numerically.
However, the Wilson fermion induces a strong chiral symmetry
breaking and thus it is important to see if this introduces
any new aspect into the problem. The suggestion in 
Ref.\cite{fujikawa2} is 
based on making {\em all} the Feynman diagrams finite, namely,
the cancellation of divergences among Feynman diagrams is not 
sufficient in general. If one applies this criterion to the 
present problem, we encounter one-loop level divergences
in some of the individual Feynman diagrams for correlation
functions  though those 
divergences cancel among bosonic and fermionic contributions.
(In disconnected vacuum diagrams, two-loop diagrams contain
divergences.)
In particular, the Wilson fermion introduces a linear divergence
to individual tadpole diagrams in $d=2$, which is absent in the 
Ginsparg-Wilson fermion.
The presence of linear divergences suggests that the 
lattice regularization is not arbitrary but need to 
be ``well-behaved'' one,  which  ensures the 
precise cancellation of these linear divergences among diagrams
for $a\neq 0$, such as in a formulation with precise lattice 
supersymmetry in the free part of the Lagrangian\footnote{The 
actual numerical analysis, for example, is simplest in the 
simplest form of the Lagrangian, but due care is required to 
ensure supersymmetry in the limit $a\rightarrow 0$.}.
If all the Feynman diagrams should be absolutely convergent, the 
latticization would enjoy more freedom to recover 
supersymmetry in the limit $a\rightarrow 0$.

The second issue analyzed is the role played by extra couplings 
introduced by an analysis of  Nicolai mapping\cite{nicolai}. The 
Nicolai mapping in the present context suggests an appearance of
 exta interactions\cite{sakai}\cite{beccaria}
\cite{catterall}which vanish if the Leibniz rule is satisfied 
on the lattice. 
Also these extra terms spoil the naive hypercubic symmetry on 
the lattice. Because of these novel features of the extra
interactions, one may hope that these extra terms might remedy 
the failure of the Leibniz rule appearing in the remaining 
conventional interaction terms. We analyze this issue in the 
framework of perturbation theory. Our result shows that 
these extra terms do not completely remedy supersymmetry in 
correlation functions, which is 
broken by the failure of the Leibniz rule, at least in weak 
coupling perturbation theory, though those terms ensure the 
vanishing of vacuum energy arising from disconnected diagrams.
  As far as the correlation functions are concerned,  
supersymmetry is recovered in the limit $a\rightarrow 0$ 
{\em with or without} those extra couplings.

\section{N=2 Wess-Zumino model and Nicolai mapping}

We start with a Lagrangian defined in terms of the Wilson
fermion on the $d=2$ Euclidean lattice
\begin{eqnarray}
{\cal L}&=&\bar{\psi}(D_{(1)}+D_{(2)})\psi 
- m\bar{\psi}\psi
- 2g\bar{\psi}(P_{+}\phi P_{+}
+P_{-}\phi^{\star}P_{-})\psi\nonumber\\
&-&\phi^{\star}D_{(1)}^{\dagger}D_{(1)}\phi + F^{\star}F
-m[F\phi+(F\phi)^{\star}]
-g[F\phi^{2}+(F\phi^{2})^{\star}]\nonumber\\
&+& FD_{(2)}\phi + (FD_{(2)}\phi)^{\star}\nonumber\\
&+& g\phi^{2}(\nabla_{1}^{S}+i\nabla_{2}^{S})\phi
+g(\phi^{2}(\nabla_{1}^{S}+i\nabla_{2}^{S})\phi)^{\star}
\end{eqnarray}
where $\psi$ is a two-dimensional Dirac spinor. 
Since we are interested in the $d=2$ model as a toy model for 
4-dimensional theory, we choose the superpotential to be a 
specific form
\begin{equation}
W^{\prime}(\phi)=m\phi+g\phi^{2}.
\end{equation}

Here we defined 
\begin{eqnarray}
D_{(1)}\psi(x)&\equiv&\gamma^{\mu}\nabla^{S}_{\mu}\psi(x)
=\gamma^{\mu}\frac{1}{2a}(\psi(x+a\hat{\mu})-\psi(x-a\hat{\mu}))
,\nonumber\\
D_{(2)}\psi(x)&\equiv&\sum_{\mu}\nabla^{A}_{\mu}\psi(x)
\nonumber\\
&=&\sum_{\mu}
\frac{1}{2a}(\psi(x+a\hat{\mu})+\psi(x-a\hat{\mu})-2\psi(x)).
\end{eqnarray}
We note the important property 
$\sum_{x}f(x)(\nabla^{S}_{\mu}g)(x)
=-\sum_{x}(\nabla^{S}_{\mu}f)(x)g(x)$.
Our Euclidean $\gamma$ matrix convention is 
\begin{eqnarray}
&&(\gamma^{\mu})^{\dagger}=\gamma^{\mu},\nonumber\\
&&\gamma_{5}^{\dagger}=\gamma_{5},\nonumber\\
&&P_{\pm}=\frac{1}{2}(1\pm\gamma_{5}).
\end{eqnarray}
 When we have the operator 
$D_{(1)}^{\dagger}D_{(1)}$ in the bosonic sector, we adopt 
the convention to discard the $2\times2$ unit matrix.
The terms 
\begin{equation}
{\cal L}_{kin}=
\bar{\psi}D_{(1)}\psi 
-\phi^{\star}D^{\dagger}_{(1)}D_{(1)}\phi
+F^{\star}F
\end{equation}
stand for the kinetic (Kahler) terms.
The last two 
terms in (2.1) are the extra terms introduced by an argument of 
Nicolai mapping\cite{sakai}, while other terms are the naive 
lattice translation of the continuum theory except for the Wilson
term and its super-partners. 
Namely, the terms 
\begin{equation}
{\cal L}_{W}=
\bar{\psi}D_{(2)}\psi+FD_{(2)}\phi+(FD_{(2)}\phi)^{\star}
\end{equation}
stand for a naive supersymmetrization of the Wilson term, which 
 induce a hard breaking of continuum chiral symmetry.

The last two terms in (2.1) vanish if the 
 Leibniz rule for $\nabla_{\mu}^{S}$ should hold on the 
lattice, namely, if 
$\phi^{2}(x)(\nabla_{\mu}^{S}\phi)(x)
=\frac{1}{3}(\nabla_{\mu}^{S}\phi^{3})(x)$.
This fact suggests that the extra terms might remedy the 
supersymmetry breaking induced by the failure of the Leibniz rule
in the remaining interaction terms. 
These extra terms also break the (hyper-)cubic symmetry on the 
lattice.

The elimination of the auxiliary fields $F$ and $F^{\star}$
in the starting Lagrangian gives
\begin{eqnarray}
{\cal L}&=&\bar{\psi}(D_{(1)}+D_{(2)})\psi 
- \bar{\psi}(P_{+}W^{\prime\prime}
+P_{-}(W^{\prime\prime})^{\star})\psi\nonumber\\
&-&\phi^{\star}D_{(1)}^{\dagger}D_{(1)}\phi 
- (D_{(2)}\phi)^{\star}D_{(2)}\phi
+(W^{\prime})^{\star}D_{(2)}\phi 
+ W^{\prime}(D_{(2)}\phi)^{\star}
-(W^{\prime})^{\star}W^{\prime}\nonumber\\
&+& W^{\prime}(\nabla_{1}^{S}+i\nabla_{2}^{S})\phi
+ (W^{\prime}(\nabla_{1}^{S}+i\nabla_{2}^{S})\phi)^{\star}
\end{eqnarray}
with $W^{\prime}=m\phi+g\phi^{2}$ by noting  
\begin{eqnarray}
\sum_{x}\phi(x)\nabla^{S}_{\mu}\phi(x)=-
\sum_{x}\nabla^{S}_{\mu}\phi(x)\phi(x)=0.
\end{eqnarray}
This Lagrangian agrees with the one introduced in 
Refs.\cite{sakai}\cite{beccaria}\cite{catterall}
by an analysis of Nicolai mapping\footnote{Recently, the 
Nicolai mapping was extended to the $d=2$ Wess-Zumino model
defined in terms of Ginsparg-Wilson operators,
which makes chiral symmetry manifest\cite{kiku-naka}.}.

The Nicolai mapping here is defined by
\begin{eqnarray}
&&\xi_{1}=(\nabla^{S}_{1}+D_{(2)})A-U-\nabla^{S}_{2}B,\nonumber\\
&&\xi_{2}=(-\nabla^{S}_{1}+D_{(2)})B-V-\nabla^{S}_{2}A
\end{eqnarray}
with 
\begin{eqnarray}
&&\phi(x)=\frac{1}{\sqrt{2}}(A+iB),\nonumber\\
&&U=\frac{1}{\sqrt{2}}(W^{\prime}+(W^{\prime})^{\star}),
\nonumber\\
&&V=\frac{1}{\sqrt{2}i}(W^{\prime}-(W^{\prime})^{\star}).
\end{eqnarray}
If one uses the specific representation of $\gamma$ matrices
\begin{equation}
\gamma^{1}=\left(\begin{array}{cc}
            1&0\\
            0&-1
            \end{array}\right),
\gamma^{2}=\left(\begin{array}{cc}
            0&-1\\
            -1&0
            \end{array}\right),
\gamma_{5}=\left(\begin{array}{cc}
            0&i\\
            -i&0
            \end{array}\right)
\end{equation}
the Jacobian for the transformation from $(\xi_{1},\xi_{2})$
to (A,B) precisely agrees with the determinant of the fermion 
operator in (2.7). The bosonic part of the Lagrangian is then 
written as
\begin{equation}
\sum_{x}{\cal L}_{boson}(x)=-\sum_{x}
\frac{1}{2}[\xi^{2}_{1}(x)+\xi^{2}_{2}(x)]
\end{equation}
and the partition function is given by 
\begin{eqnarray}
Z&=&\int{\cal D}\bar{\psi}{\cal D}\psi{\cal D}A{\cal D}B
\exp[\sum_{x}{\cal L}(x)]\nonumber\\
&=&\int{\cal D}\xi_{1}{\cal D}\xi_{2}\exp\{-\sum_{x}
\frac{1}{2}[\xi^{2}_{1}(x)+\xi^{2}_{2}(x)]\}
\end{eqnarray}
if one imposes universal (periodic) boundary conditions both 
on fermionic and bosonic variables. 
The vanishing of the vacuum energy is thus ensured by the 
Nicolai mapping even for $a\neq 0$. 
The partition function 
is reduced to
\begin{equation}
Tr(-1)^{\hat{F}}\exp[-\beta\hat{H}]
\end{equation}
if the continuum limit $a\rightarrow0$ is well-defined. The 
presence of the Nicolai mapping would then  ensure the 
degeneracy of bosonic and fermionic spectra of $\hat{H}$ in the 
continuum limit. 

\section{Supersymmetry transformation}
One may define a lattice supersymmetry transformation 
parametrized by 
a constant Dirac-type Grassmann parameter $\bar{\epsilon}$ by
\begin{eqnarray}
&&\delta\bar{\psi}=
-\bar{\epsilon}[P_{-}\phi+P_{+}\phi^{\star}]D_{(1)}
-\bar{\epsilon}[P_{-}F^{\star}+P_{+}F]\nonumber\\
&&\delta\psi=0\nonumber\\
&&\delta\phi=\bar{\epsilon}P_{+}\psi, \ \ \ \delta\phi^{\star}
=\bar{\epsilon}P_{-}\psi\nonumber\\
&&\delta F=\bar{\epsilon}P_{-}D_{(1)}\psi, 
\ \ \ \delta F^{\star}
=\bar{\epsilon}P_{+}D_{(1)}\psi.
\end{eqnarray}
The supersymmetry transformation parametrized by a constant 
Dirac-type Grassmann parameter $\epsilon$, which is treated to
be independent of $\bar{\epsilon}$, is given by
\begin{eqnarray}
&&\delta\psi=
-D_{(1)}[P_{-}\phi+P_{+}\phi^{\star}]\epsilon
-[P_{-}F^{\star}+P_{+}F]\epsilon\nonumber\\
&&\delta\bar{\psi}=0\nonumber\\
&&\delta\phi=\bar{\psi}P_{+}\epsilon, \ \ \ \delta\phi^{\star}
=\bar{\psi}P_{-}\epsilon\nonumber\\
&&\delta F=\bar{\psi}D_{(1)}P_{-}\epsilon, 
\ \ \ \delta F^{\star}
=\bar{\psi}D_{(1)}P_{+}\epsilon.
\end{eqnarray}

Here we note that
\begin{eqnarray}
&&D^{\dagger}_{(1)}=-D_{(1)},\nonumber\\
&&D_{(1)}\gamma_{5}+\gamma_{5}D_{(1)}=0,\nonumber\\
&&D^{\dagger}_{(2)}=D_{(2)},\nonumber\\
&&D_{(2)}\gamma_{5}-\gamma_{5}D_{(2)}=0.
\end{eqnarray}
Under the transformation (3.1), it can be confirmed that the 
kinetic term
\begin{eqnarray}
\int\delta{{\cal L}}_{kin}&=&
\int\{-\bar{\epsilon}[P_{-}\phi+P_{+}\phi^{\star}]D_{(1)}
-\bar{\epsilon}[P_{-}F^{\star}+P_{+}F]\}D_{(1)}\psi\nonumber\\
&&+\int\phi^{\star}\partial_{\mu}\partial_{\mu}
\bar{\epsilon}P_{+}\psi
+\int\phi\partial_{\mu}\partial_{\mu}
\bar{\epsilon}P_{-}\psi\nonumber\\
&&+\int F^{\star}\bar{\epsilon}P_{-}D_{(1)}\psi
+\int F\bar{\epsilon}P_{+}D_{(1)}\psi=0
\end{eqnarray}
is in fact invariant. 
The mass term
\begin{equation}
{{\cal L}}_{mass}=-m\bar{\psi}\psi-mF\phi
-mF^{\star}\phi^{\star} 
\end{equation}
and the Wilson term
\begin{eqnarray}
{{\cal L}}_{W}&=&
\bar{\psi}D_{(2)}\psi+FD_{(2)}\phi+(FD_{(2)}\phi)^{\star}
\end{eqnarray}
are also confirmed to be invariant under the above supersymmetry 
transformation by  using the relation
$D_{(1)}D_{(2)}=D_{(2)}D_{(1)}$.

The variation of the (conventional) interaction terms
\begin{equation}
{{\cal L}}_{int}
=- 2g\bar{\psi}(P_{+}\phi P_{+}
+P_{-}\phi^{\star}P_{-})\psi
-g[F\phi^{2}+(F\phi^{2})^{\star}]
\end{equation}
is given by 
\begin{eqnarray}
\int\delta{{\cal L}}_{int}
&=&g\int\bar{\epsilon}[-2P_{-}(D_{(1)}\phi)\phi\psi-
2P_{+}(D_{(1)}\phi^{\star})\phi^{\star}\psi\nonumber\\
&&+P_{-}(D_{(1)}\phi^{2})\psi
+P_{+}(D_{(1)}(\phi^{\star})^{2})\psi]
\end{eqnarray}
where we used the relation $\sum_{x}f(x)(D_{(1)}g)(x)=
-\sum_{x}(D_{(1)}f)(x)g(x)$.
This variation of the interaction terms would vanish if the 
difference operator $D_{(1)}$ should satisfy the Leibniz rule
$2(D_{(1)}\phi)(x)\phi(x)=(D_{(1)}\phi^{2})(x)$.
(The terms quadratic in $\psi$ vanish by themselves.)

As for the $U(1)\times U_{R}(1)$ charges (and holomorphicity)
analogous to those in 
the 4-dimensional theory\cite{seiberg}, we may assign
\begin{eqnarray}
&&\phi = (1,1),\nonumber\\
&&F = (1 ,-1),\nonumber\\
&&P_{+}\psi = (1,0),\ \ \ P_{-}\psi = (-1,0), \nonumber\\
&&m = (-2,0),\nonumber\\
&&g = (-3,-1)
\end{eqnarray}
for the terms appearing in the conventional formulation.
Even for $m=g=0$, the Wilson term ${{\cal L}}_{W}$ in the 
Lagrangian violates the $U(1)$ symmetry.

As for the extra terms introduced by an argument of 
Nicolai mapping
\begin{eqnarray}
{\cal L}_{extra}
&=&g\phi^{2}(\nabla_{1}^{S}+i\nabla_{2}^{S})\phi
+(g\phi^{2}(\nabla_{1}^{S}+i\nabla_{2}^{S})\phi)^{\star}
\end{eqnarray}
which break hypercubic symmetry, they  are not 
invariant under the above supersymmetry transformation.
Since the lattice supersymmetry transformation we defined above
preserves transformation  properties under the hypercubic 
symmetry, the supersymmetry variation of these extra terms do 
not mix with the variation of other terms. Those extra terms 
also break $U_{R}(1)$ symmetry.
We re-iterate that these extra terms vanish if the Leibniz rule
should hold on the lattice.

\section{Feynman rules for perturbative calculations}
It is interesting to examine to what extent the extra terms 
introduced by an argument of Nicolai 
mapping\cite{sakai}\cite{beccaria}\cite{catterall}, namely 
the last two terms in ${\cal L}$ (2.1), are essential to maintain
supersymmetry in perturbation theory.
Although the extra terms, which break hypercubic symmetry,
 do not mix with other terms under the 
lattice supsersymmetry transformation in (3.1) as we already 
explained,  the lattice supersymmetry transformation is not 
unique and thus we cannot {\em a priori} exclude the possible 
cancellation of supersymmetry breaking effects among the 
interaction terms\footnote{In fact, an exact Ward identity
which is regarded as a part of supersymmetry is known to exist
in the Lagrangian defined by the Nicolai 
mapping\cite{beccaria}\cite{catterall}. We shall analyze this 
identity later.}. 
Our assumption is that perturbative 
calculations are universal at least for a small coupling 
 constant and 
independent of the specific definitions of lattice supersymmetry 
transformation.
One starts with the free part of the Lagrangian
\begin{eqnarray}
{\cal L}_{0}&=&\bar{\psi}(D_{(1)}+D_{(2)})\psi 
-\phi^{\star}D_{(1)}^{\dagger}D_{(1)}\phi + F^{\star}F
- m\bar{\psi}\psi - m[F\phi+(F\phi)^{\star}]\nonumber\\
&+& FD_{(2)}\phi + (FD_{(2)}\phi)^{\star}.
\end{eqnarray}
 When we have the operator 
$D_{(1)}^{\dagger}D_{(1)}$ in the bosonic sector, we adopt 
the convention to discard the $2\times2$ unit matrix.

The propagators are given by
\begin{eqnarray}
\langle\phi\phi^{\star}\rangle&=&\frac{1}
{D_{(1)}^{\dagger}D_{(1)}+(-m+D_{(2)})^{2}},\nonumber\\
\langle F F^{\star}\rangle&=&(-)
\frac{D_{(1)}^{\dagger}D_{(1)}}
{D_{(1)}^{\dagger}D_{(1)}+(-m+D_{(2)})^{2}},\nonumber\\
\langle F \phi\rangle
&=&\langle F^{\dagger} \phi^{\dagger}\rangle
=(-)\frac{(-m+D_{(2)})}
{D_{(1)}^{\dagger}D_{(1)}+(-m+D_{(2)})^{2}},\nonumber\\
\langle\psi\bar{\psi}\rangle
&=&\frac{-1}{D_{(1)}+D_{(2)}-m}
=(-)
\frac{-D_{(1)}+D_{(2)}-m}
{D_{(1)}^{\dagger}D_{(1)}+(-m+D_{(2)})^{2}}
\end{eqnarray}
and other propagators vanish.   Here we have
$D_{(1)}^{\dagger}=-D_{(1)}$.
In the momentum representation, we have
\begin{eqnarray}
\langle\phi\phi^{\star}\rangle&=&\frac{a^{2}}
{(\sin ak_{\mu})^{2}+(a{\cal M})^{2}},\nonumber\\
\langle F F^{\star}\rangle&=&(-)
\frac{(\sin ak_{\mu})^{2}}
{(\sin ak_{\mu})^{2}+(a{\cal M})^{2}},\nonumber\\
\langle F \phi\rangle
&=&\langle F^{\star} \phi^{\star}\rangle
=\frac{a(a{\cal M})}
{(\sin ak_{\mu})^{2}+(a{\cal M})^{2}},\nonumber\\
\langle\psi\bar{\psi}\rangle
&=&
\frac{ai\gamma^{\mu}\sin ak_{\mu}+a(a{\cal M})}
{(\sin ak_{\mu})^{2}+(a{\cal M})^{2}}
\end{eqnarray}
where
\begin{equation}
{\cal M}(ak_{\mu})\equiv \sum_{\mu}\frac{1}{a}(1-\cos ak_{\mu})
+m.
\end{equation}

The interaction terms for perturbative calculations are 
given by
\begin{eqnarray}
{\cal L}_{int}&=&
- 2g\bar{\psi}(P_{+}\phi P_{+}
+P_{-}\phi^{\star}P_{-})\psi
-g[F\phi^{2}+(F\phi^{2})^{\star}]\nonumber\\
&+& g\phi^{2}(\nabla_{1}^{S}+i\nabla_{2}^{S})\phi
+ g(\phi^{2}(\nabla_{1}^{S}+i\nabla_{2}^{S})\phi)^{\star}.
\end{eqnarray}
 If one 
sets one of $\phi$ in the extra interaction terms (i.e., in 
the last two terms in ${\cal L}_{int}$) to be a constant
\begin{equation}
\phi(x)=\phi_{0}= constant
\end{equation}
then the extra terms go away by noting 
$\sum_{x}\phi(x)\nabla^{S}_{\mu}\phi(x)=0$.  This 
means that 
the effects of the extra terms introduced by the Nicolai mapping
do not appear in the one-loop
diagrams if one sets the momenta of external lines at $0$.
In other words, only those diagrams where the momenta of 
external lines cannot be set to be zero are affected by those 
extra terms.  

\section{Lower order diagrams}
We now examine several lower order diagrams for correlation 
functions in perturbation
theory\footnote{A perturbative analysis of the 4-dimensional 
Wess-Zumino model with the Wilson fermion was performed in 
\cite{bartels}. In the present $d=2$ model, the perturbation is 
based on $g/m\ll 1$.}. The theory in $d=2$ becomes more 
convergent 
in higher order diagrams, and one can confirm that the possible 
supersymmetry breaking effects in higher order diagrams become 
irrelevant in the limit $a\rightarrow0$, provided that one-loop
sub-diagrams are properly treated. The one-loop 
diagrams are thus crucial in the analysis of supersymmetry.
As for the disconnected vacuum diagrams, they shall be later
analyzed separately. 

\subsection{Tadpole diagrams}
The one-loop tadpole diagrams for the scalar $\phi$ consist 
of two diagrams; the first one is a scalar loop and the 
second is a fermion loop. 
The scalar loop contribution is given by
\begin{eqnarray}
-2g\phi\langle F\phi \rangle&=&
-2g\phi\int^{\pi/a}_{-\pi/a}\frac{d^{2}k}{(2\pi)^{2}}
\frac{a(a{\cal M})}
{(\sin ak_{\mu})^{2}+(a{\cal M})^{2}}\nonumber\\
&=&-2g\phi\int^{\pi}_{-\pi}\frac{d^{2}k}{(2\pi)^{2}}(\frac{1}{a})
\frac{(a{\cal M}(k_{\mu}))}
{(\sin k_{\mu})^{2}+(a{\cal M}(k_{\mu}))^{2}}
\end{eqnarray}
where we re-scaled the integration variable in the last 
expression as 
\begin{equation}
ak_{\mu}\rightarrow k_{\mu}
\end{equation}
and defined 
\begin{equation}
a{\cal M}(k_{\mu})=\sum_{\mu}(1-\cos k_{\mu})
+am.
\end{equation}
In the limit $a\rightarrow0$ this diagram diverges as 
$\sim 1/a$, namely, linearly divergent. This divergence is 
worse than the divergence in continuum theory (and also in the 
lattice theory with the Ginsparg-Wilson 
fermion\footnote{The power counting with the Ginsparg-Wilson
fermion is identical to that of continuum 
theory\cite{fuji-ishi}.}), which is 
logarithmic. 

The fermion loop contribution is given by 
\begin{eqnarray}
2g\phi Tr P_{+}\langle\psi\bar{\psi}\rangle
&=&2g\phi\int^{\pi/a}_{-\pi/a}\frac{d^{2}k}{(2\pi)^{2}}
Tr P_{+}\frac{ai\gamma^{\mu}\sin ak_{\mu}+a(a{\cal M})}
{(\sin ak_{\mu})^{2}+(a{\cal M})^{2}}\nonumber\\
&=&
2g\phi\int^{\pi}_{-\pi}\frac{d^{2}k}{(2\pi)^{2}}(\frac{1}{a})
\frac{(a{\cal M}(k_{\mu}))}
{(\sin k_{\mu})^{2}+(a{\cal M}(k_{\mu}))^{2}}
\end{eqnarray}
which is precisely cancelled by the scalar contribution (5.1) 
for a finite $a$. However, each diagram is  linearly 
divergent due to the strong chiral symmetry breaking by the 
Wilson term. In a numerical simulation,  
one would need to choose the free part of the Lagrangian to be
lattice supersymmetric, as in the present formulation, so that 
the cancellation of linear divergence is 
exact\footnote{If all the Feynman diagrams should be absolutely 
convergent, one would enjoy 
more freedom in choosing lattice regularization. I thank 
H. Kawai and T. Onogi for a helpful comment on this point.}.
Alternatively, one may introduce an auxiliary 
regularization such as higher derivative regularization 
to make each diagram convergent and thus less sensitive to the 
parameter $a$ in the limit $a\rightarrow0$, as we discussed in 
4-dimensional theory\cite{fujikawa2}.

\subsection{Induced $\phi^{2}$-coupling}
We have contributions from a scalar loop and a fermion loop.
The scalar loop contribution is given by
\begin{eqnarray}
&&\frac{1}{2!}(2g)^{2}\phi\langle F\phi\rangle
\langle F\phi\rangle\phi\nonumber\\
&=&2g^{2}\phi^{2}\int^{\pi/a}_{-\pi/a}
\frac{d^{2}k}{(2\pi)^{2}}\frac{a(a{\cal M})}
{\sin^{2}(ak_{\mu}+ap_{\mu})+(a{\cal M})^{2}}\frac{a(a{\cal M})}
{(\sin ak_{\mu})^{2}+(a{\cal M})^{2}}\nonumber\\
&=&2g^{2}\phi(-p_{\mu})\phi(p_{\mu})
\int^{\pi}_{-\pi}\frac{d^{2}k}{(2\pi)^{2}}
\frac{(a{\cal M}(k_{\mu}+ap_{\mu}))}
{\sin^{2}(ak_{\mu}+ap_{\mu})+(a{\cal M}(k_{\mu}+ap_{\mu}))^{2}}
\nonumber\\
&&\qquad \ \ \ \ \ \ \ \ \times\frac{(a{\cal M}(k_{\mu}))}
{(\sin k_{\mu})^{2}+(a{\cal M}(k_{\mu}))^{2}}
\end{eqnarray}
which approaches a constant for $a\rightarrow0$. This behavior
is consistent with the continuum behavior, but the difference is
that all the momentum regions, not only the infrared region,
contribute to the integral. This is an effect of the chiral
symmetry breaking by the Wilson term.

The fermion loop contribution gives
\begin{eqnarray}
&&\frac{-1}{2!}(2g)^{2}Tr\phi
\langle P_{+}\psi\bar{\psi}P_{+}\rangle
\langle P_{+}\psi\bar{\psi}P_{+}\rangle\phi\nonumber\\
&=&-2g^{2}\phi(-p_{\mu})\phi(p_{\mu})
\int^{\pi}_{-\pi}\frac{d^{2}k}{(2\pi)^{2}}
\frac{(a{\cal M}(k_{\mu}+ap_{\mu}))}
{\sin^{2}(k_{\mu}+ap_{\mu})+(a{\cal M}(k_{\mu}+ap_{\mu}))^{2}}
\nonumber\\
&&\qquad \ \ \ \ \ \ \ \ \times\frac{(a{\cal M}(k_{\mu}))}
{(\sin k_{\mu})^{2}+(a{\cal M}(k_{\mu}))^{2}}
\end{eqnarray}
which is precisely cancelled by the scalar contribution (5.5)
even for a finite lattice spacing $a$.

Each Feynman diagram which gives induced couplings higher powers 
in $\phi$ such as $\phi^{3}$ is reduced to the continuum result
in the limit $a\rightarrow0$. Those couplings in any case cancel
 among the scalar and fermion contributions even for finite 
$a$. The non-renormalization of the superpotential in this 
sense is thus maintained in the one-loop level, and in 
higher-loop levels in the limit $a\rightarrow 0$.

\subsection{Self-energy corrections}
The simplest self-energy correction is that to the auxiliary
fields $F$ and $F^{\star}$. The one-loop correction is
given by 
\begin{eqnarray}
&&\frac{1}{2!}2g^{2}F\langle\phi^{2}(\phi^{\star})^{2}\rangle 
F^{\star}\nonumber\\
&=&2g^{2}FF^{\star}\langle\phi\phi^{\star}\rangle
\langle\phi\phi^{\star}\rangle\nonumber\\
&=&2g^{2}FF^{\star}\int^{\pi/a}_{-\pi/a}
\frac{d^{2}k}{(2\pi)^{2}}
\frac{a^{2}}
{\sin^{2}(ak_{\mu}+ap_{\mu})+(a{\cal M})^{2}}
\frac{a^{2}}
{(\sin ak_{\mu})^{2}+(a{\cal M})^{2}}\nonumber\\
&=&2g^{2}F(p_{\mu})F^{\star}(p_{\mu})\int^{\pi}_{-\pi}
\frac{d^{2}k}{(2\pi)^{2}}
\frac{1}{\sin^{2}k_{\mu}+(a{\cal M}(k_{\mu}))^{2}}
\nonumber\\
&&\qquad\times
\frac{a^{2}}
{\sin^{2}(k_{\mu}+ap_{\mu})+(a{\cal M}(k_{\mu}+ap_{\mu}))^{2}}.
\end{eqnarray}
This integral vanishes for $a\rightarrow0$ if one keeps 
the integration domain {\em outside} 
\begin{equation}
|k_{\mu}|< \delta \ \ \ for \ \ \ all \ \ \ \mu
\end{equation}
for arbitrarily small but finite $\delta$ and for fixed 
$p_{\mu}$. The integration 
inside the above domain gives a finite continuum result if 
one notes
\begin{equation}
(\sin k_{\mu})^{2}+(a{\cal M}(k_{\mu}))^{2}\simeq
k^{2}_{\mu}+(\frac{1}{2}k^{2}_{\mu}+am)^{2}\simeq
k^{2}_{\mu}+ amk^{2}_{\mu}+(am)^{2}
\end{equation}
inside the above domain. A re-scaling of $k_{\mu}$ back to 
the original momentum variables $k_{\mu}\rightarrow ak_{\mu}$ 
gives the continuum result in the limit $a\rightarrow0$. 
\\

The fermion self-energy correction is given by
\begin{eqnarray}
&&\frac{1}{2!}
(2g)^{2}\bar{\psi}(P_{+}\phi P_{+}
+P_{-}\phi^{\star}P_{-})\psi\bar{\psi}(P_{+}\phi P_{+}
+P_{-}\phi^{\star}P_{-})\psi 
\nonumber\\
&\rightarrow&(2g)^{2}[\bar{\psi}(P_{+}\phi P_{+}
\langle\psi\bar{\psi}\rangle P_{-}\phi^{\star}P_{-})\psi  
+\bar{\psi}(P_{-}\phi^{\star}P_{-})\langle\psi\bar{\psi}\rangle
(P_{+}\phi P_{+})\psi]\nonumber\\
&\rightarrow&(2g)^{2}[\bar{\psi}P_{+}
\langle\psi\bar{\psi}\rangle P_{-}\psi
\langle\phi\phi^{\star}\rangle  
+\bar{\psi}P_{-}\langle\psi\bar{\psi}\rangle P_{+}\psi
\langle\phi\phi^{\star}\rangle]
\\
&=&(2g)^{2}\bar{\psi}(p_{\mu})\int^{\pi/a}_{-\pi/a}
\frac{d^{2}k}{(2\pi)^{2}}[\frac{ai\gamma^{\mu}\sin(ap_{\mu}
+ak_{\mu})}
{\sin^{2}(ap_{\mu}+ak_{\mu})+(a{\cal M})^{2}}\frac{a^{2}}
{(\sin ak_{\mu})^{2}+(a{\cal M})^{2}}]\psi(p_{\mu})\nonumber\\
&=&(2g)^{2}\bar{\psi}(p_{\mu})\int^{\pi}_{-\pi}
\frac{d^{2}k}{(2\pi)^{2}}[\frac{ai\gamma^{\mu}\sin(ap_{\mu}
+k_{\mu})}
{\sin^{2}(ap_{\mu}+k_{\mu})+(a{\cal M})^{2}}\frac{1}
{(\sin k_{\mu})^{2}+(a{\cal M})^{2}}]\psi(p_{\mu})\nonumber.
\end{eqnarray}
This integral vanishes if one sets $p_{\mu}=0$, which means that 
the fermion mass receives no quantum correction when 
renormalized at vanishing momentum, despite the chiral symmetry 
breaking by the Wilson term\footnote{This vanishing mass 
correction arises from the differences of the Feynman rules in 
the present model and QCD. Also, all the higher loop 
corrections are reduced to the (supersymmetric) continuum 
results in the limit $a\rightarrow 0$ in the present model.
The Wilson term does not always imply the mass shift.}. 
This integral also vanishes for the 
domain outside (5.8) in the limit $a\rightarrow 0$, and the 
integral is reduced to the continuum result for the domain 
inside (5.8) in the limit $a\rightarrow 0$ for fixed $p_{\mu}$.  
 
To analyze the wave function renormalization, we consider the 
case with an infinitesimal $p_{\mu}$ but the lattice spacing 
$a$ kept fixed\footnote{It should be noted that we assume a small
coupling $g/m\ll 1$ and infinitesimally small external momentum 
$p_{\mu}$ but otherwise make no assumtion about the lattice 
spacing $a$. }.  
Namely,
\begin{equation}
|p_{\mu}|\ll m,\ \  1/a.
\end{equation}
We thus expand
\begin{eqnarray}
\sin(ap_{\mu}+k_{\mu})&&\simeq \sin k_{\mu}
+ap_{\mu} \cos k_{\mu},
\nonumber\\
a{\cal M}(k_{\mu}+ap_{\mu})&&\simeq a{\cal M}(k_{\mu})
+ ap_{\mu}\sin k_{\mu},\nonumber\\
\sin^{2}(ap_{\mu}+k_{\mu})+(a{\cal M})^{2}&&\simeq 
\sin^{2}(k_{\mu})+(a{\cal M}(k_{\mu}))^{2}\nonumber\\
&&+ap_{\mu}\sin 2k_{\mu}+2a{\cal M}(k_{\mu})ap_{\mu}\sin k_{\mu}.
\end{eqnarray}
 
The integral in this expression is given by
\begin{eqnarray}
a^{2}\int^{\pi}_{-\pi}
\frac{d^{2}k}{(2\pi)^{2}}&&
\{\frac{\sum_{\mu}i\gamma^{\mu}p_{\mu} \cos k_{\mu}}
{[\sin^{2}(k_{\mu})+(a{\cal M}(k))^{2}]^{2}}\nonumber\\
&&-\frac{\sum_{\mu}i\gamma^{\mu}\sin k_{\mu}[p_{\nu}\sin 2k_{\nu}
+2(a{\cal M}(k))p_{\nu}\sin k_{\nu}]}
{[\sin^{2}(k_{\mu})+(a{\cal M}(k))^{2}]^{3}}\}\nonumber\\
=a^{2}\int^{\pi}_{-\pi}
\frac{d^{2}k}{(2\pi)^{2}}&&
\{\frac{\sum_{\mu}i\gamma^{\mu}p_{\mu} \cos k_{\mu}}
{[\sin^{2}(k_{\mu})+(a{\cal M}(k))^{2}]^{2}}\nonumber\\
&&+\frac{1}{2}\sum_{\mu}i\gamma^{\mu}\sin k_{\mu}
\sum_{\nu}p_{\nu}\frac{\partial}{\partial k_{\nu}}
\frac{1}{[\sin^{2}(k_{\mu})+(a{\cal M}(k))^{2}]^{2}}\}\nonumber\\
=\frac{1}{2}a^{2}\int^{\pi}_{-\pi}
\frac{d^{2}k}{(2\pi)^{2}}&&
\{\frac{\sum_{\mu}i\gamma^{\mu}p_{\mu} \cos k_{\mu}}
{[\sin^{2}(k_{\mu})+(a{\cal M}(k))^{2}]^{2}}\}.
\end{eqnarray}
By noting the symmetry under 
$k_{1}\leftrightarrow k_{2}$,
we thus have the wave function renormalization for the fermion
\begin{eqnarray}
&&2g^{2}\int^{\pi}_{-\pi}
\frac{d^{2}k}{(2\pi)^{2}}[\frac{a^{2}\frac{1}{2}
\sum_{\nu} \cos k_{\nu}}
{[\sin^{2}(k_{\mu})+(a{\cal M}(k))^{2}]^{2}}]
\bar{\psi}(p_{\mu})i\gamma^{\mu}p_{\mu}\psi(p_{\mu})
\end{eqnarray}
which disagrees with the finite renormalization factor for the 
fields $F$ and $F^{\star}$ at $p_{\mu}=0$ in (5.7) for a 
finite $a$; 
\begin{eqnarray}
&&\int^{\pi}_{-\pi}
\frac{d^{2}k}{(2\pi)^{2}}
\frac{a^{2}}
{[(\sin k_{\mu})^{2}+(a{\cal M}(k_{\mu}))^{2}]^{2}}
-\int^{\pi}_{-\pi}
\frac{d^{2}k}{(2\pi)^{2}}\frac{a^{2}\frac{1}{2}
\sum_{\nu} \cos k_{\nu}}
{[\sin^{2}(k_{\mu})+(a{\cal M}(k))^{2}]^{2}}
\nonumber\\
&&=\int^{\pi}_{-\pi}
\frac{d^{2}k}{(2\pi)^{2}}
\frac{a^{2}\frac{1}{2}
\sum_{\nu}(1-\cos k_{\nu})}
{[(\sin k_{\mu})^{2}+(a{\cal M}(k_{\mu}))^{2}]^{2}}> 0
\end{eqnarray}
for $a\neq 0$, though this difference vanishes in the limit 
$a\rightarrow 0$. This shows that the finite wave function 
renormalization factor {\em breaks} supersymmetry for $a\neq 0$.

We next examine the self-energy corrections to the scalar 
field $\phi$. The contribution from a scalar loop diagram in
the {\em conventional} interaction terms gives 
\begin{eqnarray}
&&\frac{1}{2!}g^{2}[F\phi^{2}+(F\phi^{2})^{\star}]
[F\phi^{2}+(F\phi^{2})^{\star}]
\nonumber\\
&&\rightarrow g^{2}\langle(F\phi^{2})^{\star}F\phi^{2}\rangle
\nonumber\\
&&=4g^{2}\phi^{\star}\langle F^{\star}F\rangle
\langle\phi^{\star}\phi\rangle\phi\\
&&=-4g^{2}\phi^{\star}(p_{\mu})\phi(p_{\mu})\int^{\pi}_{-\pi}
\frac{d^{2}k}{(2\pi)^{2}}\nonumber\\
&&\times
\frac{(\sin k_{\mu})^{2}}
{(\sin k_{\mu})^{2}+(a{\cal M})^{2}}
\frac{1}
{\sin^{2}(k_{\mu}+ap_{\mu})+(a{\cal M}(k_{\mu}+ap_{\mu}))^{2}}.
\nonumber
\end{eqnarray}
The one-loop fermion contribution is given by
\begin{eqnarray}
&&\frac{1}{2!}(2g)^{2}\bar{\psi}(P_{+}\phi P_{+}
+P_{-}\phi^{\star}P_{-})\psi\bar{\psi}(P_{+}\phi P_{+}
+P_{-}\phi^{\star}P_{-})\psi
\nonumber\\
&&\rightarrow
(2g)^{2}\bar{\psi}P_{-}\phi^{\star}P_{-}\psi
\bar{\psi}P_{+}\phi P_{+}\psi
\nonumber\\
&&\rightarrow
-(2g)^{2}\phi^{\star}
Tr[\langle P_{+}\psi\bar{\psi}P_{-}\rangle
\langle P_{-}\psi\bar{\psi}P_{+}\rangle]\phi
\nonumber\\
&&=-(2g)^{2}\phi^{\star}\frac{1}{2}\int^{\pi}_{-\pi}
\frac{d^{2}k}{(2\pi)^{2}}
\nonumber\\
&&\times Tr[\frac{i\gamma^{\mu}\sin(k_{\mu}+ap_{\mu})}
{\sin^{2}(k_{\mu}+ap_{\mu})+(a{\cal M}(k_{\mu}+ap_{\mu}))^{2}}
\frac{i\gamma^{\mu}\sin k_{\mu}}
{(\sin k_{\mu})^{2}+(a{\cal M})^{2}}]\phi
\nonumber\\
&&=4g^{2}\phi^{\star}(p_{\mu})\phi(p_{\mu})
\int^{\pi}_{-\pi}\frac{d^{2}k}{(2\pi)^{2}}
\nonumber\\
&&\times[\frac{\sin(k_{\mu}+ap_{\mu})\sin k_{\mu}}
{\sin^{2}(k_{\mu}+ap_{\mu})+(a{\cal M}(k_{\mu}+ap_{\mu}))^{2}}
\frac{1}
{(\sin k_{\mu})^{2}+(a{\cal M})^{2}}].
\end{eqnarray}
The sum of these two contributions gives rise to
\begin{eqnarray}
&&4g^{2}\phi^{\star}(p_{\mu})\phi(p_{\mu})
\int^{\pi}_{-\pi}\frac{d^{2}k}{(2\pi)^{2}}
\nonumber\\
&&\times[\frac{(\sin(k_{\mu}+ap_{\mu})-\sin k_{\mu})
\sin k_{\mu}}
{\sin^{2}(k_{\mu}+ap_{\mu})+(a{\cal M}(k_{\mu}+ap_{\mu}))^{2}}
\frac{1}
{(\sin k_{\mu})^{2}+(a{\cal M})^{2}}]
\end{eqnarray}
which vanishes for $p_{\mu}=0$. This means that the mass 
correction to the scalar particles exactly vanishes in the 
one-loop level. However, each term logarithmically diverges 
in the limit $a\rightarrow0$, which suggests that the choice of
the free-part of the Lagrangian should be at least invariant 
under the lattice supersymmetry transformation to ensure
the divergence cancellation, such as in the present formulation.
This integral vanishes for $a\rightarrow 0$ for the domain 
outside (5.8) and for fixed $p_{\mu}$. For the the domain 
inside (5.8) and for fixed $p_{\mu}$, the integral is reduced to
 the continuum result in the limit $a\rightarrow 0$.

For an infinitesimal $p_{\mu}$, we have
\begin{eqnarray}
&&4g^{2}\phi^{\star}(p_{\mu})\phi(p_{\mu})
\int^{\pi}_{-\pi}\frac{d^{2}k}{(2\pi)^{2}}\{
\frac{\sum_{\mu}ap_{\mu}\cos k_{\mu}\sin k_{\mu}}
{(\sin k_{\mu})^{2}+(a{\cal M})^{2}}
\nonumber\\
&&\times\sum_{\nu}ap_{\nu}\frac{\partial}{\partial k_{\nu}}
[\frac{1}
{\sin^{2}(k_{\mu})+(a{\cal M}(k_{\mu}))^{2}}]
-\frac{\frac{1}{2}\sum_{\mu}[(ap_{\mu})^{2}\sin^{2}k_{\mu}]}
{[(\sin k_{\mu})^{2}+(a{\cal M})^{2}]^{2}}\}
\nonumber\\
&&=4g^{2}\phi^{\star}(p_{\mu})\phi(p_{\mu})
\int^{\pi}_{-\pi}\frac{d^{2}k}{(2\pi)^{2}}\{
\sum_{\mu}ap_{\mu}\cos k_{\mu}\sin k_{\mu}
\nonumber\\
&&\frac{1}{2}\times\sum_{\nu}ap_{\nu}
\frac{\partial}{\partial k_{\nu}}
\frac{1}
{[\sin^{2}(k_{\mu})+(a{\cal M}(k_{\mu}))^{2}]^{2}}
-\frac{\frac{1}{2}\sum_{\mu}[(ap_{\mu})^{2}\sin^{2}k_{\mu}]}
{[(\sin k_{\mu})^{2}+(a{\cal M})^{2}]^{2}}\}
\nonumber\\
&&=-2g^{2}\phi^{\star}(p_{\mu})\phi(p_{\mu})
\int^{\pi}_{-\pi}\frac{d^{2}k}{(2\pi)^{2}}
a^{2}\sum_{\mu}[p^{2}_{\mu}\cos 2k_{\mu}
+p^{2}_{\mu}\sin^{2}k_{\mu}]
\nonumber\\
&&\times
\frac{1}
{[\sin^{2}(k_{\mu})+(a{\cal M}(k_{\mu}))^{2}]^{2}}
\nonumber\\
&&=-2g^{2}\phi^{\star}(p_{\mu})p^{2}_{\mu}\phi(p_{\mu})
\int^{\pi}_{-\pi}\frac{d^{2}k}{(2\pi)^{2}}
a^{2}\frac{1}{2}\sum_{\mu}[\cos 2k_{\mu}
+\frac{1}{2}(1-\cos 2k_{\mu})]
\nonumber\\
&&\times
\frac{1}
{[\sin^{2}(k_{\mu})+(a{\cal M}(k_{\mu}))^{2}]^{2}}.
\end{eqnarray}
This deviates from the renormalization of $F$ and $F^{\star}$
for finite $a$, 
\begin{eqnarray}
&&\int^{\pi}_{-\pi}
\frac{d^{2}k}{(2\pi)^{2}}
\frac{a^{2}}
{[\sin^{2} k_{\mu}+(a{\cal M}(k_{\mu}))^{2}]^{2}}\nonumber\\
&&
-\int^{\pi}_{-\pi}
\frac{d^{2}k}{(2\pi)^{2}}\frac{a^{2}\frac{1}{2}
\sum_{\nu}[\cos 2k_{\nu}+\frac{1}{2}(1-\cos 2k_{\mu})]}
{[\sin^{2}k_{\mu}+(a{\cal M}(k))^{2}]^{2}}
\nonumber\\
&&=\int^{\pi}_{-\pi}
\frac{d^{2}k}{(2\pi)^{2}}
\frac{a^{2}\frac{1}{4}
\sum_{\nu}(1-\cos 2k_{\nu})}
{[\sin^{2} k_{\mu}+(a{\cal M}(k_{\mu}))^{2}]^{2}}> 0
\end{eqnarray}
though this difference vanishes in the limit $a\rightarrow0$.

\subsection{Self-energy corrections induced by extra couplings}

Finally, we examine the effects of the exta couplings in (4.5)
introduced by an argument of Nicolai mapping  on the self-energy
of scalar particles. This is given by
\begin{eqnarray}
&&\frac{1}{2!}[g\phi^{2}(\nabla_{1}^{S}+i\nabla_{2}^{S})\phi
+ g(\phi^{2}(\nabla_{1}^{S}+i\nabla_{2}^{S})\phi)^{\star}]
\nonumber\\
&&\times [g\phi^{2}(\nabla_{1}^{S}+i\nabla_{2}^{S})\phi
+ g(\phi^{2}(\nabla_{1}^{S}+i\nabla_{2}^{S})\phi)^{\star}]
\nonumber\\
&&\rightarrow
g^{2}
[(\phi^{2}(\nabla_{1}^{S}+i\nabla_{2}^{S})\phi)^{\star}]
[\phi^{2}(\nabla_{1}^{S}+i\nabla_{2}^{S})\phi]
\nonumber\\
&&\rightarrow 
2g^{2}((\nabla_{1}^{S}+i\nabla_{2}^{S})\phi)^{\star}
\langle\phi^{\star}\phi\rangle\langle\phi^{\star}\phi\rangle
(\nabla_{1}^{S}+i\nabla_{2}^{S})\phi
\nonumber\\
&&+4g^{2}((\nabla_{1}^{S}+i\nabla_{2}^{S})\phi)^{\star}
\langle\phi^{\star}\phi\rangle\langle
\phi^{\star}(\nabla_{1}^{S}+i\nabla_{2}^{S})\phi\rangle
\phi
\nonumber\\
&&+4g^{2}\phi^{\star}
\langle((\nabla_{1}^{S}+i\nabla_{2}^{S})\phi)^{\star}\phi
\rangle
\langle\phi^{\star}\phi\rangle
(\nabla_{1}^{S}+i\nabla_{2}^{S})\phi
\nonumber\\
&&+4g^{2}\phi^{\star}
\langle((\nabla_{1}^{S}+i\nabla_{2}^{S})\phi)^{\star}\phi
\rangle
\langle\phi^{\star}(\nabla_{1}^{S}+i\nabla_{2}^{S})\phi\rangle
\phi
\nonumber\\
&&+4g^{2}\phi^{\star}
\langle((\nabla_{1}^{S}+i\nabla_{2}^{S})\phi)^{\star}
(\nabla_{1}^{S}+i\nabla_{2}^{S})\phi\rangle
\langle\phi^{\star}\phi\rangle
\phi.
\end{eqnarray}
The first term in (5.21) gives
\begin{eqnarray}
&&2g^{2}((\nabla_{1}^{S}+i\nabla_{2}^{S})\phi(p_{\mu}))^{\star}
(\nabla_{1}^{S}+i\nabla_{2}^{S})\phi(p_{\mu})
\int^{\pi}_{-\pi}
\frac{d^{2}k}{(2\pi)^{2}}
\nonumber\\
&&\times
\frac{1}
{\sin^{2}(k_{\mu}+ap_{\mu})+(a{\cal M}(k_{\mu}+ap_{\mu}))^{2}}
\frac{a^{2}}
{\sin^{2}(k_{\mu})+(a{\cal M}(k_{\mu}))^{2}}
\nonumber\\
&&=2g^{2}\phi(p_{\mu})^{\star}(\frac{\sin ap_{\mu}}{a})^{2}
\phi(p_{\mu})
\int^{\pi}_{-\pi}
\frac{d^{2}k}{(2\pi)^{2}}
\nonumber\\
&&\times
\frac{1}
{\sin^{2}(k_{\mu}+ap_{\mu})+(a{\cal M}(k_{\mu}+ap_{\mu}))^{2}}
\frac{a^{2}}
{\sin^{2}(k_{\mu})+(a{\cal M}(k_{\mu}))^{2}}.
\end{eqnarray}
This gives for an infinitesimal $p_{\mu}$
\begin{eqnarray}
2g^{2}\phi(p_{\mu})^{\star}p^{2}_{\mu}\phi(p_{\mu})
\int^{\pi}_{-\pi}
\frac{d^{2}k}{(2\pi)^{2}}
\frac{a^{2}}
{[\sin^{2}(k_{\mu})+(a{\cal M}(k_{\mu}))^{2}]^{2}}.
\end{eqnarray}
The second term gives
\begin{eqnarray}
&&4g^{2}((i\frac{\sin ap_{1}}{a}-\frac{\sin ap_{2}}{a})
\phi(p_{\mu}))^{\star}\phi(p_{\mu})
\int^{\pi}_{-\pi}
\frac{d^{2}k}{(2\pi)^{2}}
\nonumber\\
&&\times
\frac{a}
{\sin^{2}(k_{\mu})+(a{\cal M}(k_{\mu}))^{2}}
\frac{-i\sin(k_{1}+ap_{1})
+\sin(k_{2}+ap_{2})}
{\sin^{2}(k_{\mu}+ap_{\mu})+(a{\cal M}(k_{\mu}+ap_{\mu}))^{2}}
\end{eqnarray}
which gives for an infinitesimal $p_{\mu}$
\begin{eqnarray}
&&4g^{2}(-ip_{1}-p_{2})\phi(p_{\mu})^{\star}\phi(p_{\mu})
\int^{\pi}_{-\pi}
\frac{d^{2}k}{(2\pi)^{2}}
\frac{a}
{\sin^{2}(k_{\mu})+(a{\cal M}(k_{\mu}))^{2}}
\nonumber\\
&&\times
[\frac{-iap_{1}\cos k_{1}
+ap_{2}\cos k_{2}}
{\sin^{2}(k_{\mu})+(a{\cal M}(k_{\mu}))^{2}}
\nonumber\\
&&+(-i\sin k_{1}+\sin k_{2})\sum_{\nu}ap_{\nu}\frac{\partial}
{\partial k_{\nu}}
\frac{1}{\sin^{2}(k_{\mu})+(a{\cal M}(k_{\mu}))^{2}}]
\nonumber\\
&&=4g^{2}(-ip_{1}-p_{2})\phi(p_{\mu})^{\star}\phi(p_{\mu})
\int^{\pi}_{-\pi}
\frac{d^{2}k}{(2\pi)^{2}}
\frac{a}
{\sin^{2}(k_{\mu})+(a{\cal M}(k_{\mu}))^{2}}
\nonumber\\
&&\times\frac{1}{2}
[\frac{-iap_{1}\cos k_{1}
+ap_{2}\cos k_{2}}
{\sin^{2}(k_{\mu})+(a{\cal M}(k_{\mu}))^{2}}]
\nonumber\\
&&=g^{2}(-ip_{1}-p_{2})\phi(p_{\mu})^{\star}\phi(p_{\mu})
\int^{\pi}_{-\pi}
\frac{d^{2}k}{(2\pi)^{2}}
\frac{a}
{\sin^{2}(k_{\mu})+(a{\cal M}(k_{\mu}))^{2}}
\nonumber\\
&&\times
[\frac{(-iap_{1}+ap_{2})\sum_{\nu}\cos k_{\nu}}
{\sin^{2}(k_{\mu})+(a{\cal M}(k_{\mu}))^{2}}]
\nonumber\\
&&=g^{2}(-p^{2}_{\mu})\phi(p_{\mu})^{\star}\phi(p_{\mu})
\int^{\pi}_{-\pi}
\frac{d^{2}k}{(2\pi)^{2}}
\frac{a^{2}\sum_{\nu}\cos k_{\nu}}
{[\sin^{2}(k_{\mu})+(a{\cal M}(k_{\mu}))^{2}]^{2}}.
\end{eqnarray}
The third term gives the complex conjugate of the second 
term, which agrees with the second term itself. Thus we have
altogether
\begin{eqnarray}
-2g^{2}\phi(p_{\mu})^{\star}p^{2}_{\mu}\phi(p_{\mu})
\int^{\pi}_{-\pi}
\frac{d^{2}k}{(2\pi)^{2}}
\frac{a^{2}\sum_{\nu}\cos k_{\nu}}
{[\sin^{2}(k_{\mu})+(a{\cal M}(k_{\mu}))^{2}]^{2}}.
\end{eqnarray}
The fourth and the fifth terms give precisely the negative of 
the contributions of the conventional interactions to the 
self-energy of scalar particles, (5.16) and (5.17), which we 
have already evaluated.

If we collect all the terms arising from the extra couplings
together, we obtain
\begin{eqnarray}
2g^{2}\phi(p_{\mu})^{\star}p^{2}_{\mu}\phi(p_{\mu})
\int^{\pi}_{-\pi}
\frac{d^{2}k}{(2\pi)^{2}}
\frac{a^{2}[1-\sum_{\nu}\cos k_{\nu}
+\frac{1}{2}\sum_{\nu}\cos2k_{\nu}
+\frac{1}{4}\sum_{\nu}(1-\cos2k_{\nu})]}
{[\sin^{2}(k_{\mu})+(a{\cal M}(k_{\mu}))^{2}]^{2}}
\end{eqnarray}
which vanishes in the limit $a\rightarrow0$, as it should be
since the conventional interaction already ensures supersymmetry
in the limit $a\rightarrow0$. These terms do not help 
the wave function renormalization factor of $\phi$  agree with 
that of either $F$ or $\psi$. The breaking of supersymmetry in 
the wave function renormalization factors persist for $a\neq 0$
even if one includes the effects of the extra couplings induced 
by an analysis of Nicolai mapping.
\\

\section{Low-energy effective action with one-loop corrections}

The low-energy effective action which includes the 
one-loop quantum corrections is written in a momentum 
representation as
\begin{eqnarray}
{\cal L}_{eff}&=&(1+z_{\psi})\bar{\psi}i\pslash\psi 
-m\bar{\psi}\psi
-(1+z_{\phi})\phi^{\star}p_{\mu}^{2}\phi 
-\frac{m^{2}}{1+z_{F}}\phi^{\star}\phi+...\nonumber\\
&=&(1+z_{\psi})[\bar{\psi}i\pslash\psi 
-\frac{m}{1+z_{\psi}}\bar{\psi}\psi]\nonumber\\
&-&(1+z_{\phi})[\phi^{\star}p_{\mu}^{2}\phi 
+\frac{m^{2}}{(1+z_{F})(1+z_{\phi})}\phi^{\star}\phi]+...
\end{eqnarray}
after the elimination of the auxiliary fields $F$ and 
$F^{\star}$. Here we defined the finite wave function 
renormalization factors (see (5.7), (5.14) and (5.19))
\begin{eqnarray}
&&z_{F}=2g^{2}\int^{\pi}_{-\pi}
\frac{d^{2}k}{(2\pi)^{2}}
\frac{a^{2}}
{[(\sin k_{\mu})^{2}+(a{\cal M}(k_{\mu}))^{2}]^{2}},
\nonumber\\
&&z_{\psi}=2g^{2}\int^{\pi}_{-\pi}
\frac{d^{2}k}{(2\pi)^{2}}[\frac{a^{2}\frac{1}{2}
\sum_{\nu} \cos k_{\nu}}
{[\sin^{2}(k_{\mu})+(a{\cal M}(k))^{2}]^{2}}],\nonumber\\
&&z_{\phi}=2g^{2}\int^{\pi}_{-\pi}
\frac{d^{2}k}{(2\pi)^{2}}[\frac{a^{2}[\frac{1}{2}
\sum_{\nu} \cos 2k_{\nu}+\frac{1}{4}\sum_{\nu}(1-\cos 2k_{\nu})}
{[\sin^{2}(k_{\mu})+(a{\cal M}(k))^{2}]^{2}}].
\end{eqnarray}
Supersymmetry suggests the uniform wave function renormalization
\begin{equation}
1+z_{\psi}=1+z_{\phi}
\end{equation}
and the  degeneracy of the mass parameter
\begin{equation}
\frac{m^{2}}{(1+z_{\psi})^{2}}
=\frac{m^{2}}{(1+z_{F})(1+z_{\phi})}
\end{equation}
or in the accuracy of one-loop correction
\begin{equation}
2z_{\psi}=(z_{F}+z_{\phi}).
\end{equation}
If one includes the contributions from the extra couplings,
these conditions are replaced by
\begin{eqnarray}
&&z_{\psi}=z_{\phi}+z_{extra},\nonumber\\
&&2z_{\psi}=(z_{F}+z_{\phi}+z_{extra})
\end{eqnarray}
with (see (5.27))
\begin{eqnarray}
z_{extra}&=&2g^{2}\int^{\pi}_{-\pi}
\frac{d^{2}k}{(2\pi)^{2}}
\frac{a^{2}[-1+\sum_{\nu}\cos k_{\nu}
-\frac{1}{2}\sum_{\nu}\cos2k_{\nu}
-\frac{1}{4}\sum_{\nu}(1-\cos2k_{\nu})]}
{[\sin^{2}(k_{\mu})+(a{\cal M}(k_{\mu}))^{2}]^{2}}\nonumber\\
&=&2z_{\psi}-z_{F}-z_{\phi}.
\end{eqnarray}
It is interesting that the degeneracy of the mass parameter,
namely, the second condition in (6.6) is satisfied even for 
finite $a$ in the presence of the extra couplings. However,
the uniform wave function renormalization condition
\begin{equation}
z_{\psi}=z_{F}
\end{equation}
is still broken since $z_{\psi}<z_{F}$ for finite $a$.
The supersymmetry is thus broken for $a\neq 0$ even with the 
extra couplings induced by the Nicolai mapping.

In the continuum limit $a\rightarrow0$, we have
\begin{eqnarray}
&&z_{\psi}=z_{F}=z_{\phi},\nonumber\\
&&z_{extra}=0
\end{eqnarray}
and the supersymmetry is recovered, {\em with or without} the 
extra couplings. This conclusion is valid up to any finite order
in perturbation theory.

\section{Check of Ward identity}
The Nicolai mapping suggests that the Lagrangian is written 
as
\begin{equation}
{\cal L}=\bar{\psi}^{\alpha}(x)\frac{\partial \xi_{\alpha}(x)}
{\partial A^{\beta}(y)}\psi^{\beta}(y)
-\frac{1}{2}\sum_{\alpha}(\xi_{\alpha}(x))^{2}
\end{equation}
where\footnote{We identify the spinor index of the Dirac fermion 
with the flavor index of the scalar particle, an apparently 
Lorentz non-invariant operation.} 
\begin{equation}
\{A_{\alpha}\}=(A,B),
\end{equation}
and thus we have the relation
\begin{equation}
-\langle\psi^{\alpha}(x)\bar{\psi}^{\beta}(y)\rangle
=\int{\cal D}\xi 
\frac{\partial A_{\alpha}(x)}{\partial\xi_{\beta}(y)}
\exp[-\sum_{x}\frac{1}{2}\sum_{\alpha}(\xi_{\alpha}(x))^{2}]
\end{equation}
which is equal to 
\begin{equation}
\langle A_{\alpha}(x)\xi_{\beta}(y)\rangle
=\int{\cal D}\xi 
\frac{\partial A_{\alpha}(x)}{\partial\xi_{\beta}(y)}
\exp[-\sum_{x}\frac{1}{2}\sum_{\alpha}(\xi_{\alpha}(x))^{2}]
\end{equation} 
as can be confirmed by expanding $A_{\alpha}(x)$ formally 
in powers of $\xi_{\kappa}(z)$.
These relations give rise to the identity\cite{beccaria}
\cite{catterall} 
\begin{equation}
\langle\psi^{\alpha}(x)\bar{\psi}^{\beta}(y)\rangle
+ \langle A_{\alpha}(x)\xi_{\beta}(y)\rangle=0.
\end{equation}

We check this identity for a small momentum region.
The fermion propagator with one-loop quantum corrections is
given by
\begin{equation}
\langle\psi\bar{\psi}\rangle=\frac{1}{-i(1+z_{\psi})\pslash+m}
=\frac{1}{m}+\frac{i(1+z_{\psi})\pslash}{m^{2}}+ O(p_{\mu}^{2})
\end{equation}
in the low-energy limit $|\pslash/m|\ll 1$ but with fixed $a$.

We next note
\begin{eqnarray}
&&U=mA+\frac{g}{\sqrt{2}}(A^{2}-B^{2}),\nonumber\\
&&V=mB+\frac{2g}{\sqrt{2}}AB.
\end{eqnarray}
We evaluate
\begin{eqnarray}
\langle\xi_{1}(x)A(y)\rangle&=&
(\nabla^{S}_{1}+D_{(2)})\langle A(x)A(y)\rangle
-m\langle A(x)A(y)\rangle\nonumber\\
&&-\frac{g}{\sqrt{2}}
\langle(A^{2}-B^{2})(x)A(y)\rangle
-\nabla^{S}_{2}\langle B(x)A(y)\rangle
\end{eqnarray}
with the interaction terms
\begin{eqnarray}
{\cal L}_{int}&=&-\frac{g}{\sqrt{2}}((m-D_{(2)})A)(A^{2}-B^{2})
-\frac{2g}{\sqrt{2}}((m-D_{(2)})B)(AB)
\nonumber\\
&&-\frac{g}{4}[(A^{2}-B^{2})^{2}+4(AB)^{2}]
\nonumber\\
&&
+\frac{g}{\sqrt{2}}(\nabla^{S}_{1}A-\nabla^{S}_{2}B)(A^{2}-B^{2})
+\frac{2g}{\sqrt{2}}(-\nabla^{S}_{1}B-\nabla^{S}_{2}A)(AB)
\nonumber\\
&&-\frac{2g}{\sqrt{2}}\bar{\psi}(A+iB\gamma_{5})\psi.
\end{eqnarray}
We first have
\begin{eqnarray}
&&(\nabla^{S}_{1}+D_{(2)})\langle A(x)A(y)\rangle
-m\langle A(x)A(y)\rangle\nonumber\\
&&=
\frac{-ip_{1}-m}{(1+z_{\phi})p^{2}_{\mu}+\frac{m^{2}}{(1+z_{F})}}
\nonumber\\
&&=-\frac{(1+z_{F})}{m}-\frac{ip_{1}(1+z_{F})}{m^{2}}
+O(p^{2}_{\mu}).
\end{eqnarray}
We next evaluate
\begin{equation}
-\frac{g}{\sqrt{2}}
\langle(A^{2}-B^{2})(x)A(y)\rangle.
\end{equation}
The contributions from the 
conventional interaction terms give
in momentum representation
\begin{eqnarray}
&&2g^{2}\{\int^{\pi}_{-\pi}
\frac{d^{2}k}{(2\pi)^{2}}
\frac{a^{2}}{(\sin k_{\mu})^{2}+(a{\cal M}(k_{\mu}))^{2}}
\frac{1}
{(\sin k_{\mu}+ap_{\mu})^{2}+(a{\cal M}(k_{\mu}+ap_{\mu}))^{2}}\}
\nonumber\\
&&\times\frac{a^{2}{\cal M}(ap_{\mu})}
{(\sin ap_{\mu})^{2}+(a{\cal M}(ap_{\mu}))^{2}} 
\nonumber\\
\end{eqnarray}
which gives for small $p_{\mu}$ 
\begin{equation}
\frac{z_{F}}{m}.
\end{equation}
The contributions from the extra couplings give 
in momentum representation
\begin{eqnarray}
&&2g^{2}\{\int^{\pi}_{-\pi}
\frac{d^{2}k}{(2\pi)^{2}}
\frac{a^{2}}{(\sin k_{\mu})^{2}+(a{\cal M}(k_{\mu}))^{2}}
\frac{1}
{(\sin k_{\mu}+ap_{\mu})^{2}+(a{\cal M}(k_{\mu}+ap_{\mu}))^{2}}\}
\nonumber\\
&&\times\frac{a i\sin ap_{1}}
{(\sin ap_{\mu})^{2}+(a{\cal M}(ap_{\mu}))^{2}} 
\nonumber\\
&&-4g^{2}\{\int^{\pi}_{-\pi}
\frac{d^{2}k}{(2\pi)^{2}}
\frac{a}{(\sin k_{\mu})^{2}+(a{\cal M}(k_{\mu}))^{2}}
\frac{i(\sin k_{1}+ap_{1})}
{(\sin k_{\mu}+ap_{\mu})^{2}+(a{\cal M}(k_{\mu}+ap_{\mu}))^{2}}\}
\nonumber\\
&&\times\frac{a^{2}}
{(\sin ap_{\mu})^{2}+(a{\cal M}(ap_{\mu}))^{2}} 
\end{eqnarray}
give for small $p_{\mu}$
\begin{equation}
(z_{F}-z_{\psi})\frac{ip_{1}}{m^{2}}.
\end{equation}
This term vanishes for $a\rightarrow0$.

These calculations show that the Ward identity\cite{beccaria}
\cite{catterall} 
\begin{equation}
\langle\psi\bar{\psi}\rangle_{11}+\langle\xi_{1}(x)A(y)\rangle
=0
\end{equation}
is precisely satisfied up to 
the order $O(p^{2}_{\mu})$,
\begin{eqnarray}
\langle\psi\bar{\psi}\rangle&=&\frac{1}{-i(1+z_{\psi})\pslash+m}
=\frac{1}{m}+\frac{i(1+z_{\psi})\pslash}{m^{2}}
+ O(p_{\mu}^{2}),
\nonumber\\
\langle\xi_{1}(x)A(y)\rangle&=&
(\nabla^{S}_{1}+D_{(2)})\langle A(x)A(y)\rangle
-m\langle A(x)A(y)\rangle\nonumber\\
&&\qquad -\frac{g}{\sqrt{2}}
\langle(A^{2}-B^{2})(x)A(y)\rangle
-\nabla^{S}_{2}\langle B(x)A(y)\rangle
\nonumber\\
&=&-\frac{(1+z_{F})}{m}-\frac{ip_{1}(1+z_{F})}{m^{2}}
+\frac{z_{F}}{m}+(z_{F}-z_{\psi})\frac{ip_{1}}{m^{2}}
+O(p^{2}_{\mu}),\nonumber\\
&=&-\frac{1}{m}-\frac{i(1+z_{\psi})p_{1}}{m^{2}}
+ O(p_{\mu}^{2})
\end{eqnarray}
even for $z_{F}\neq z_{\psi}$ at $a\neq 0$, if one recalls
$(\pslash)_{11}=p_{1}$. But in any case 
the correction terms induced by the extra interactions vanish 
,$z_{F}-z_{\psi}\rightarrow0$, in the limit $a\rightarrow0$. 
Consequently, the Ward identity in the limit $a\rightarrow0$ is 
not modified by the extra interactions introduced by the Nicolai 
mapping.
This is consistent with the numerical findings of the behavior 
of various exact and not-exact Ward identities, which appear
to be equally valid numerically in the limit $a\rightarrow0$
\cite{catterall}. 

\section{Disconnected vacuum diagrams}
Contributions to the vacuum energy cancel exactly for the free 
part of the 
Lagrangian due to the precise lattice supersymmetry.
The lowest non-trivial contributions from interaction terms 
arise in the two-loop level. The (conventional) two-loop scalar
contribution is given by
\begin{eqnarray}
&&\frac{1}{2!}g^{2}[F\phi^{2}+(F\phi^{2})^{\star}]
[F\phi^{2}+(F\phi^{2})^{\star}]
\nonumber\\
&&\rightarrow g^{2}\langle(F\phi^{2})^{\star}F\phi^{2}\rangle
\nonumber\\
&&=2g^{2}\langle F^{\star}F\rangle
\langle\phi^{\star}\phi\rangle\langle\phi^{\star}\phi\rangle
\\
&&=-2g^{2}\int^{\pi}_{-\pi}
\frac{d^{2}p}{(2\pi)^{2}}\int^{\pi}_{-\pi}
\frac{d^{2}k}{(2\pi)^{2}}\frac{1}
{(\sin p_{\mu})^{2}+(a{\cal M}(p_{\mu}))^{2}}
\nonumber\\
&&\times
\frac{(\sin k_{\mu})^{2}}
{(\sin k_{\mu})^{2}+(a{\cal M}(k_{\mu}))^{2}}
\frac{1}
{\sin^{2}(k_{\mu}+p_{\mu})+(a{\cal M}(k_{\mu}+p_{\mu}))^{2}}.
\nonumber
\end{eqnarray}
The two-loop contribution which contains a fermion loop is 
given by
\begin{eqnarray}
&&\frac{1}{2!}(2g)^{2}\bar{\psi}(P_{+}\phi P_{+}
+P_{-}\phi^{\star}P_{-})\psi\bar{\psi}(P_{+}\phi P_{+}
+P_{-}\phi^{\star}P_{-})\psi
\nonumber\\
&&\rightarrow
(2g)^{2}\bar{\psi}P_{-}\phi^{\star}P_{-}\psi
\bar{\psi}P_{+}\phi P_{+}\psi
\nonumber\\
&&\rightarrow
-(2g)^{2}\langle\phi^{\star}\phi\rangle
Tr[\langle P_{+}\psi\bar{\psi}P_{-}\rangle
\langle P_{-}\psi\bar{\psi}P_{+}\rangle]
\nonumber\\
&&=-(2g)^{2}\frac{1}{2}\int^{\pi}_{-\pi}
\frac{d^{2}p}{(2\pi)^{2}}\int^{\pi}_{-\pi}
\frac{d^{2}k}{(2\pi)^{2}}\frac{1}
{(\sin p_{\mu})^{2}+(a{\cal M}(p_{\mu}))^{2}}
\nonumber\\
&&\times Tr[\frac{i\gamma^{\mu}\sin(k_{\mu}+p_{\mu})}
{\sin^{2}(k_{\mu}+p_{\mu})+(a{\cal M}(k_{\mu}+p_{\mu}))^{2}}
\frac{i\gamma^{\mu}\sin k_{\mu}}
{(\sin k_{\mu})^{2}+(a{\cal M}(p_{\mu}))^{2}}]\phi
\nonumber\\
&&=4g^{2}\int^{\pi}_{-\pi}
\frac{d^{2}p}{(2\pi)^{2}}
\int^{\pi}_{-\pi}\frac{d^{2}k}{(2\pi)^{2}}
\frac{1}
{(\sin p_{\mu})^{2}+(a{\cal M}(p_{\mu}))^{2}}
\nonumber\\
&&\times[\frac{\sin(k_{\mu}+p_{\mu})\sin k_{\mu}}
{\sin^{2}(k_{\mu}+p_{\mu})+(a{\cal M}(k_{\mu}+p_{\mu}))^{2}}
\frac{1}
{(\sin k_{\mu})^{2}+(a{\cal M}(k_{\mu}))^{2}}].
\end{eqnarray}
In the limit $a\rightarrow 0$, these two integrals contain
logarithmically divergent infrared sigularities in the re-scaled
variables $p_{\mu}$ and $k_{\mu}$. These divergences, which 
agree with the divergences in continuum theory, precisely
cancel each other. However, the remaining finite parts of 
these two integrals do not quite cancel each other even in the 
limit $a\rightarrow 0$ and thus 
lead to the non-vanishing vacuum energy. This is a result of 
supersymmetry breaking by the failure of the Leibniz rule.
This complication arises since the vacuum diagrams are not 
finite even in $d=2$ (in fact contain logarithmic 
overlap-divergence) 
and all the momentum regions contribute to these vacuum diagrams.

A way to remove these finite contributions in the limit 
$a\rightarrow 0$ (without relying on the extra couplings) is to 
apply a higher derivative 
regularization\cite{fujikawa2} which amounts to the replacement 
of all the terms in the free part of the Lagrangian (4.1) as
\begin{eqnarray}
{\cal L}_{0}&=&\bar{\psi}(D_{(1)}+D_{(2)})R\psi 
-\phi^{\star}D_{(1)}^{\dagger}D_{(1)}R\phi + F^{\star}RF
\nonumber\\
&-& m\bar{\psi}R\psi - m[FR\phi+(FR\phi)^{\star}]
+ FD_{(2)}R\phi + (FD_{(2)}R\phi)^{\star}
\end{eqnarray}
where $R$ is the higher derivative regulator
\begin{eqnarray}
R=\frac{D^{\dagger}_{(1)}D_{(1)}+ D^{2}_{(2)}+M^{2}}{M^{2}}
\end{eqnarray}
with a new fixed mass scale $M$. This regularization preserves
the lattice supersymmetry (3.1) and (3.2) in the free part of 
the Lagrangian.  By this way, all the 
non-vanishing contributions are limited to the momentum regions 
$p^{2}_{\mu}\leq M^{2}$, and the vacuum diagrams comletely 
cancel in the limit $a\rightarrow 0$.  

It is interesting to see how the extra couplings help to 
remove the vacuum energy even for $a\neq 0$.
This is given by
\begin{eqnarray}
&&\frac{1}{2!}[g\phi^{2}(\nabla_{1}^{S}+i\nabla_{2}^{S})\phi
+ g(\phi^{2}(\nabla_{1}^{S}+i\nabla_{2}^{S})\phi)^{\star}]
\nonumber\\
&&\times [g\phi^{2}(\nabla_{1}^{S}+i\nabla_{2}^{S})\phi
+ g(\phi^{2}(\nabla_{1}^{S}+i\nabla_{2}^{S})\phi)^{\star}]
\nonumber\\
&&\rightarrow
g^{2}
\langle[(\phi^{2}(\nabla_{1}^{S}+i\nabla_{2}^{S})\phi)^{\star}]
[\phi^{2}(\nabla_{1}^{S}+i\nabla_{2}^{S})\phi]\rangle
\nonumber\\
&&=4g^{2}
\langle((\nabla_{1}^{S}+i\nabla_{2}^{S})\phi)^{\star}\phi
\rangle
\langle\phi^{\star}(\nabla_{1}^{S}+i\nabla_{2}^{S})\phi\rangle
\langle\phi^{\star}\phi\rangle
\nonumber\\
&&+2g^{2}
\langle((\nabla_{1}^{S}+i\nabla_{2}^{S})\phi)^{\star}
(\nabla_{1}^{S}+i\nabla_{2}^{S})\phi\rangle
\langle\phi^{\star}\phi\rangle
\langle\phi^{\star}\phi\rangle\nonumber\\
&&=-4g^{2}\int^{\pi}_{-\pi}
\frac{d^{2}p}{(2\pi)^{2}}
\int^{\pi}_{-\pi}\frac{d^{2}k}{(2\pi)^{2}}
\frac{1}
{(\sin p_{\mu})^{2}+(a{\cal M}(p_{\mu}))^{2}}
\nonumber\\
&&\times[\frac{\sin(k_{\mu}+p_{\mu})\sin k_{\mu}}
{\sin^{2}(k_{\mu}+p_{\mu})+(a{\cal M}(k_{\mu}+p_{\mu}))^{2}}
\frac{1}
{(\sin k_{\mu})^{2}+(a{\cal M}(k_{\mu}))^{2}}]\nonumber\\
&&+2g^{2}\int^{\pi}_{-\pi}
\frac{d^{2}p}{(2\pi)^{2}}\int^{\pi}_{-\pi}
\frac{d^{2}k}{(2\pi)^{2}}\frac{1}
{(\sin p_{\mu})^{2}+(a{\cal M}(p_{\mu}))^{2}}
\nonumber\\
&&\times
\frac{(\sin k_{\mu})^{2}}
{(\sin k_{\mu})^{2}+(a{\cal M}(k_{\mu}))^{2}}
\frac{1}
{\sin^{2}(k_{\mu}+p_{\mu})+(a{\cal M}(k_{\mu}+p_{\mu}))^{2}}.
\end{eqnarray}
We thus confirm that the vacuum energies (8.1), (8.2) and
(8.5) put together completely cancel for finite $a$, and this 
is a nice aspect of the analysis based on the Nicolai mapping.

\section{Discussion and conclusion}
We have examined the $N=2$ Wess-Zumino model on $d=2$ Euclidean
lattice in connection with a restoration of the Leibniz rule 
in the limit $a\rightarrow0$. In particular, we examined the 
Wilson fermion instead of the Ginsparg-Wilson fermion. We 
also examined the effects of extra couplings introduced by an
analysis of Nicolai mapping. 

As for the Wilson fermion, it introduces linear and logarithmic 
divergences in some of individual Feynman diagrams, though those 
divergences precisely cancel among Feynman diagrams for 
correlation functions in the 
formulation which ensures supersymmetry for the free-part of the
Lagrangian. 
 In the general analysis of the Leibniz rule in 
Ref.\cite{fujikawa2}, each Feynman diagram was made finite to 
ensure the Leibniz rule in the limit $a\rightarrow0$. In such 
a case, the lattice regularization would enjoy more freedom 
since it is introduced just to allow the numerical and other 
non-perturbative analyses, and the lattice artifact is safely
removed in the limit $a\rightarrow0$.   
Each Feynman diagram in the $N=2$
Wess-Zumino model in $d=2$ which was analyzed here, however,
 is not finite in general in particular with the Wilson 
fermion, and the precise
cancellation of these divergences for {\em finite} $a$ is 
important.

As for the effects of the extra couplings introduced by an 
analysis of Nicolai mapping, which breaks hypercubic symmetry,
these couplings do not completely remedy the breaking of 
supersymmetry induced by the failure of the Leibniz rule, though 
those extra couplings ensure the vanishing vacuum energy. We 
also illustrated how the Ward identity\cite{beccaria}
\cite{catterall} is satisfied even if the uniform wave function 
renormalization, which is required by supersymmetry, is not 
satisfied for finite $a$.

For a minimal latticization of the Wess-Zumino 
model in $d=2$ which ensures lattice supersymmetry for the free
part of the Lagrangian but without the extra couplings 
\begin{eqnarray}
{\cal L}&=&\bar{\psi}(D_{(1)}+D_{(2)})\psi 
- m\bar{\psi}\psi
- 2g\bar{\psi}(P_{+}\phi P_{+}
+P_{-}\phi^{\star}P_{-})\psi\nonumber\\
&-&\phi^{\star}D_{(1)}^{\dagger}D_{(1)}\phi + F^{\star}F
-m[F\phi+(F\phi)^{\star}]
-g[F\phi^{2}+(F\phi^{2})^{\star}]\nonumber\\
&+& FD_{(2)}\phi + (FD_{(2)}\phi)^{\star},
\end{eqnarray}
we have illustrated that all the supersymmetry
breaking terms in correlation functions induced by the failure 
of the Leibniz rule are irrelevant in the sense that those 
terms all vanish in the limit $a\rightarrow0$. This is 
consistent with the general analysis of perturbatively finite 
theory on the lattice\cite{fujikawa2}.

The lattice operation implies
\begin{equation}
(\nabla (fg))(x)=(\nabla f)(x)g(x)+f(x)(\nabla g)(x)
+a(\nabla f)(x)(\nabla g)(x)
\end{equation} 
if one defines $(\nabla f)(x)=(f(x+a)-f(x))/a$, and thus the 
breaking of the Leibniz rule  is of order $O(1)$ if the momentum 
carried by field variables is of  order $O(1/a)$.    
To the extent that the derivative of field variables is 
 required in supersymmetry to balance the 
dimensionality of bosonic and fermionic variables, the Leibniz 
rule is indispensable for the validity of supersymmetry.      
It is well known that supersymmetry improves ultra-violet 
properties of field theory. In the context of lattice theory,
one may rather reverse the argument
and one may even argue that the finite theory is {\em required}
 to accommodate supersymmetry in a consistent manner 
since the conventional definition of derivative
\begin{equation} 
\frac{df(x)}{dx}=\lim_{a\rightarrow0}\frac{f(x+a)-f(x)}{a}
\end{equation}
which satisfies the Leibniz rule presumes that the momentum 
carried by the field variable
$f(x)$ is negligibly small compared to $1/a$. This is realized 
in lattice theory only if the theory is finite at least in 
perturbative sense.
    
In conclusion, our analysis is consistent with the past 
analyses of the $d=2$ Wess-Zumino model and we believe that 
our analysis gives an explanation why these past non-perturbative
numerical analyses worked\cite{beccaria}\cite{catterall}, in 
particular, both of the Ward identity which is expected to be 
exact on the lattice and those Ward identities which are 
expected to be broken by the lattice 
artifacts\footnote{I thank S. Catterall for a helpful 
communication related to this issue. }\cite{catterall}.
A numerical calculation of the mass correction also appears to 
be consistent with the (continuum) perturbative estimate, as was 
noted in \cite{catterall}.
  All the supersymmetry breaking effects in correlation functions
induced by the failure of the Leibniz rule become irrelevant in 
the limit $a\rightarrow 0$ for a finite theory. 
The existence of the Nicolai mapping in the $d=2$ Wess-Zumino
model is a nice property of a specific formulation of the
specific model, such as ensuring the vanishing vacuum energy, 
but it is
not a necessary condition for a consistent definition of  
supersymmetric  models on the lattice in the limit 
$a\rightarrow 0$.
The finiteness is a more universal condition which ensures 
supersymmetry in the limit $a\rightarrow 0$.

Finally, the analyses of other aspects of supersymmetry on 
the lattice, which were not discussed in the present paper, are 
found in Refs.\cite{curci}-\cite{kaplan}.
\\

\noindent{Note added:}

M. Golterman and D.N. Petcher\cite{golterman} analyzed related 
issues in the context of $N=1$ Wess-Zumino model in $d=2$.
I thank M. Golterman for calling the above work to my attention.
S. Elitzur and A, Schwimmer\cite{elitzur} discussed a related 
problem in a Hamiltonian formalism. I thank A. Schwimmer for 
calling their work to my attention.

\end{document}